\documentclass[final,5p,twocolumn]{elsarticle}

\usepackage{graphicx, latexsym, amssymb, amsmath, color, multirow, mathrsfs, booktabs, ifpdf}

\usepackage{amsfonts}%

\newcommand{\lrb}[1]{\left(#1\right)}
\newcommand{\lrs}[1]{\left[#1\right]}
\newcommand{\Lrb}[1]{\left\{#1\right\}}

\newcommand{\svec}[1]{{\mbox{\boldmath${ #1}$}}}

\newcommand{\RE}{{\rm E}}
\newcommand{\RD}{{\rm D}}

\newcommand{\kin}{{\text{kin.}}}
\newcommand{\RR}{{\rm R}}

\journal{Physics Letters B}

\begin{document}
\begin{frontmatter}
\title{Non-local mean field effect on nuclei near  $Z=64$ sub-shell}
\author[PEK, RIKEN]{Wen Hui Long}\ead{whlong@pku.org.cn}
\author[RIKEN]{Takashi Nakatsukasa}
\author[Aizu]{Hiroyuki Sagawa}
\author[PEK]{Jie Meng}
\author[Chiba]{Hitoshi Nakada}
\author[PEK]{Ying Zhang}
\address[PEK]{School of Physics and State Key Laboratory of Nuclear Physics and Technology, Peking University, 100871 Beijing, China}
\address[RIKEN]{Theoretical Nuclear Physics Laboratory, RIKEN Nishina Center, 351-0198 Wako, Japan}
\address[Aizu]{Center for Mathematical Sciences, University of Aizu, Aizu-Wakamatsu, 965-8580
Fukushima, Japan}
\address[Chiba]{Department of Physics, Graduate School of Science, Chiba University, Chiba 263-8522,
Japan}

\begin{abstract}
Evolutions of single-particle energies and $Z=64$ sub-shell along the isotonic chain of $N=82$ are investigated in the density dependent relativistic Hartree-Fock (DDRHF) theory in comparison with other commonly used mean field models such as Skyrme HF, Gogny HFB and density dependent relativistic Hartree model (DDRMF). The pairing is treated in the BCS scheme, except for Gogny HFB. It is pointed out that DDRHF reproduces well characteristic features of experimental $Z$-dependence of both spin orbital and pseudo-spin orbital splittings around the sub-shell closure $Z=64$. Non-local exchange terms of the isoscalar $\sigma$ and $\omega$ couplings play dominant roles in the enhancements of the spin-orbit splitting of proton $2d$ states, which is the key ingredient to give the $Z=64$ sub-shell closure properly. On the other hand, the $\pi$ and $\rho$ tensor contributions for the spin-orbit splitting cancel each other and the net effect becomes rather small. The enhancement of the sub-shell gaps towards $Z=64$ is studied by the DDRHF, for which the local terms of the scalar and vector meson couplings are found to be important.
\end{abstract}


\end{frontmatter}

During the past years the developments on the radioactive ion beam (RIB) facility have opened the
new frontier for nuclear physics -- EXOTIC NUCLEI, which have intensively challenged our
understanding on the nature of nucleus from both experimental and theoretical sides. For such
nuclear system with an extreme neutron-to-proton ratio, the systematics on the shell structure
evolution is of special significance not only for the stability itself but also for the appropriate description of its exotic modes such as the halo phenomena \cite{Tan:1985, Meng:1996}. This also
provides the new criterion for the theoretical descriptions.

As an example the sub-shell closure $Z=64$ has been identified experimentally and the increasing
shell gaps towards $Z=64$ are also found on this sub-shell closure as well as relevant spin-orbit
splitting \cite{Nagai:1981}. In addition, the high-spin isomer study along the isotonic chain of
$N=83$ also indicates such trend for the $Z=64$ sub-shell gap, which increases from 2.0MeV to
2.4MeV as the proton number increases from $Z=61$ to 65 \cite{Odahara:1997}. While within the
relativistic density functional (RDF) models, the sub-shell closure $Z=64$ is not always reproduced properly. Even as the commonly used RDF model, the relativistic mean field (RMF) theory \cite{Walecka:1974, Serot:1986} fails in describing the $Z=64$ sub-shell closure, which is strongly compressed by neighboring unphysical shell closure $Z=58$ \cite{Geng:2006, Long:2007}.

With the presence of Fock terms, the new RDF model -- density dependent relativistic Hartree-Fock
(DDRHF) theory has achieved many successes in describing the nuclear structure properties
\cite{Long:2006, Long:2006PS, Long:2007, Long:2008, Liang:2008, BYSun:2008}. Under the frame of
DDRHF, the one-pion exchange and $\rho$ tensor effects can be efficiently taken into account, which have brought significant improvements in describing shell structure and corresponding evolution
\cite{Long:2007, Long:2008}. Especially with the inclusion of tensor $\rho$, the unphysical shell
closures (58 and 92), the common disease in previous RDF calculations, have been eliminated by
DDRHF, as well as appropriate recovery of the $Z=64$ sub-shell \cite{Long:2007}. It is therefore an appropriate choice to study the enhancement concerning the $Z=64$ sub-shell structure within the
DDRHF theory.

In DDRHF, the Hamiltonian contains the degrees of freedom associated with $\sigma$ scalar, $\omega$ vector, $\rho$ vector, $\rho$ tensor, $\pi$ pseudo-vector, and photon ($A$) vector couplings \cite{Long:2007, Long:2009}. In this paper all the calculations are restricted in the spherical symmetric systems. The spherical Dirac Hartree-Fock equation for nucleon can be obtained as,
 \begin{equation}\label{Spherical}
\int d\svec r' h(\svec r, \svec r') \psi(\svec r') = \varepsilon\psi(\svec r),
 \end{equation}
where $\varepsilon$ is the single-particle energy (including the rest mass) and the single-particle Dirac Hamiltonian $ h(\svec r, \svec r')$ contains the kinetic energy $h^{\text{kin}}$, the direct and exchange part of the potential energy, respectively the local $h^{\text{D}}$ and non-local
$h^{\text{E}}$,
 \begin{subequations}\label{HamiltonianS}\begin{align}
&h^{\text{kin}}(\svec r, \svec r') = \lrs{\svec\alpha\cdot\svec p + \beta M }\delta(\svec r-\svec r'),\\
&h^{\text{D}}(\svec r, \svec r') = \lrs{\Sigma_T(\svec r)\gamma_5 + \Sigma_0(\svec
r) + \beta\Sigma_S(\svec r)}\delta(\svec r - \svec r'),\\
&h^{\text{E}}(\svec r, \svec r') = \lrb{\begin{array}{cc}Y_G(\svec r, \svec r') &Y_F(\svec r, \svec r')\\[0.5em]
X_G(\svec r, \svec r')&X_F(\svec r, \svec r')\end{array}}.
 \end{align}\end{subequations}
Among the single particle Hamiltonian (\ref{HamiltonianS}), the local self-energies $\Sigma_{S}$,
$\Sigma_{0}$ and $\Sigma_{T}$ contain the contributions from the direct (Hartree) terms
\cite{Serot:1986, Reinhard:1989, Ring:1996, Meng:2006} and the rearrangement terms
\cite{Vretenar:2005}, whereas the non-local ones $X_G$, $X_F$, $Y_G$ and $Y_F$ are contributed by
the exchange (Fock) terms \cite{Bouyssy:1987, Long:2009}. These self-energies contain effects of the occupation probabilities in the BCS scheme.

From Dirac Hartree-Fock equation (\ref{Spherical}), the contributions of the single particle energy $E = \varepsilon-M$ can be written as,
 \begin{equation}
E = E_\kin + \sum_\phi E_\phi^\RD + \sum_\phi E_\phi^\RE + E_\RR,
 \end{equation}
where $E_\kin$ corresponds to the kinetic energy, $E_\phi^\RD$ and $E_\phi^\RE$ are respectively
the direct and exchange contributions of various meson (photon) nucleon couplings, and $E_\RR$
responds for the rearrangement term.

In this work, the calculations are performed along the isotonic chain of $N=82$ by DDRHF with PKA1 \cite{Long:2007}, as compared to those by Gogny Hartree-Fock-Bogoliubov (HFB) with D1S
\cite{Berger84}, Skyrme Hartree-Fock (HF) with Sly4 \cite{Chabanat:1998}, density dependent RMF
(DDRMF) with PKDD \cite{Long04}. In the Gogny calculations, the pairing effects in open shell
nuclei are considered by Bogoliubov transformation and the same interaction (D1S) is adopted as
effective pairing force. In the other calculations, the effects of pairing correlations are treated with BCS approximation and the density dependent zero range pairing force \cite{Dobaczewski:1996} is utilized in pairing channel. The Hartree-Fock or Hartree Hamiltonian is evaluated with the occupation probabilities in the BCS approximation. In DDRHF and DDRMF calculations, the pairing strength $V_0$ is adopted as 900MeV as in Ref. \cite{Long:2007} whereas in Skyrme calculations it is set to 1100MeV. Because of the fact that the studied isotones around $Z=64$ lie in the stability valley, different pairing treatments would not bring substantial difference.

\begin{figure}[t]
\includegraphics[width=0.48\textwidth]{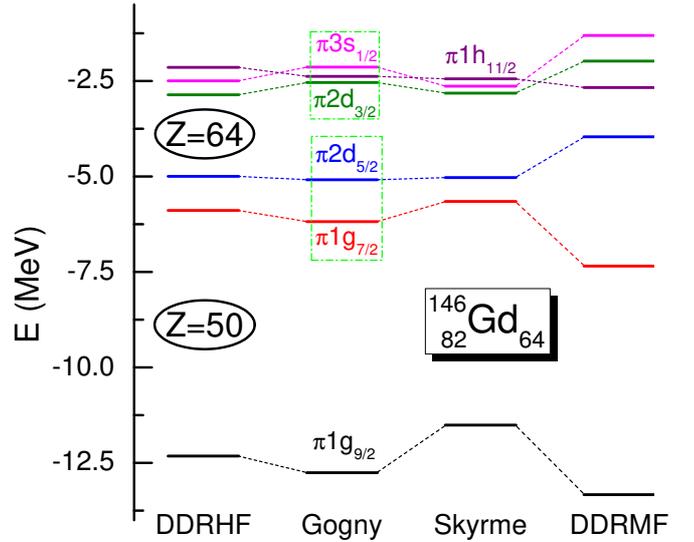}
\caption{(Color online) Proton single particle energies for $^{146}$Gd, calculated by DDRHF with
PKA1 \cite{Long:2007}, Gogny HFB with D1S \cite{Berger84}, Skyrme HF with SLy4 \cite{Chabanat:1998}
and DDRMF with PKDD \cite{Long04}.}\label{fig:LevP_Gd146}
\end{figure}

In Fig. \ref{fig:LevP_Gd146} are shown the proton ($\pi$) single particle energies calculated by DDRHF with PKA1 \cite{Long:2007}, Gogny HFB with D1S \cite{Berger84}, Skyrme HF with SLy4 \cite{Chabanat:1998} and DDRMF with PKDD \cite{Long04} for the isotone $^{146}$Gd. In this work, the Gogny HFB single particle energies are obtained by diagonalizing the HF Hamiltonian $h$ in the HFB equation, which is constructed with the density matrix of the HFB solution. Besides the magic shell ($Z=50$), DDRHF, Gogny and Skyrme models present distinct gap between the $\pi2d_{3/2}$ and $\pi2d_{5/2}$ states, namely the sub-shell closure ($Z=64$). In Fig. \ref{fig:LevP_Gd146} it seems that the recovery of $Z=64$ sub-shell is tightly related to the restoration of pseudo-spin
symmetry. Above and below the sub-shell there are two pseudo-spin doublets, respectively
$\pi2\tilde p$ ($\Lrb{\pi3s_{1/2},\pi2d_{3/2}}$) and $\pi1\tilde f$ ($\Lrb{\pi2d_{5/2},
\pi1g_{7/2}}$). Consistent with the sub-shell recovery, well conserved pseudo-spin symmetry on the doublets $\pi2\tilde p$ and $\pi1\tilde f$ are found in these three calculations. In contrast, an unphysically large gap between the $\pi2d_{5/2}$ and $\pi1g_{7/2}$ states is obtained by the DDRMF calculation, which suppresses the physical sub-shell structure $Z=64$.

\begin{figure}[t]
\includegraphics[width=0.48\textwidth]{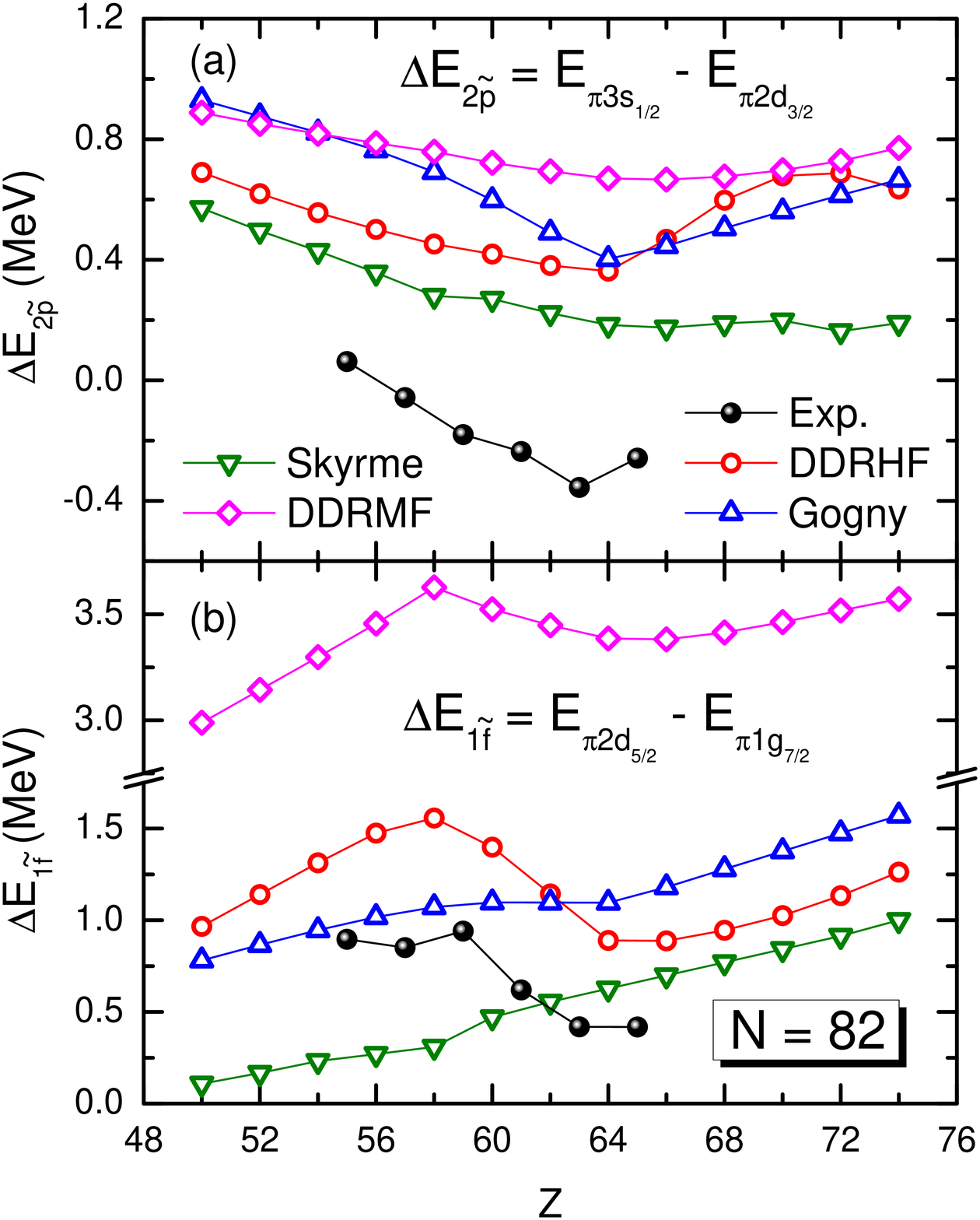}
\caption{(Color online) Pseudo-spin orbital splittings of pseudo-spin partners $\Lrb{\pi3s_{1/2},
\pi2d_{3/2}}$ (a) and $\Lrb{\pi2d_{5/2}, \pi1g_{7/2}}$ (b) along the isotonic chain of $N=82$. The
results are calculated by DDRHF with PKA1 \cite{Long:2007}, Gogny HFB with D1S \cite{Berger84},
Skyrme HF with SLy4 \cite{Chabanat:1998} and DDRMF with PKDD \cite{Long04}, in comparison with
experimental data \cite{Nagai:1981}.}\label{fig:Z64-PSS}
\end{figure}

To have a systematic understanding on that, Fig. \ref{fig:Z64-PSS} shows the pseudo-spin orbital
splittings of $\pi2\tilde p$ (Fig. \ref{fig:Z64-PSS}a) and $\pi1\tilde f$ (Fig. \ref{fig:Z64-PSS}b) as functions of proton number for $N=82$ isotones. The results are calculated by DDRHF with PKA1,
Gogny with D1S, Skyrme with Sly4 and DDRMF with PKDD. As the reference the data extracted from Ref. \cite{Nagai:1981} are also shown in Fig. \ref{fig:Z64-PSS}. Along the isotonic chain, all the
models present approximate degeneracy of the pseudo-spin doublet $\pi2\tilde p$ less than 1MeV
energy difference, while on the $\pi1\tilde f$ doublet the DDRMF calculations show a large
violation of the pseudo-spin symmetry by the amount of 3MeV, which turns out to create the
unphysical shell closure $Z=58$ \cite{Geng:2006, Long:2007}. As compared to the experimental
isotonic dependence, both DDRHF and Gogny calculation show a kink at Z=64 for the doublet
$\pi2\tilde p$, whereas for $\pi1\tilde f$ the kink is found only in DDRHF calculations with PKA1.

\begin{figure}[t]
\includegraphics[width=0.48\textwidth]{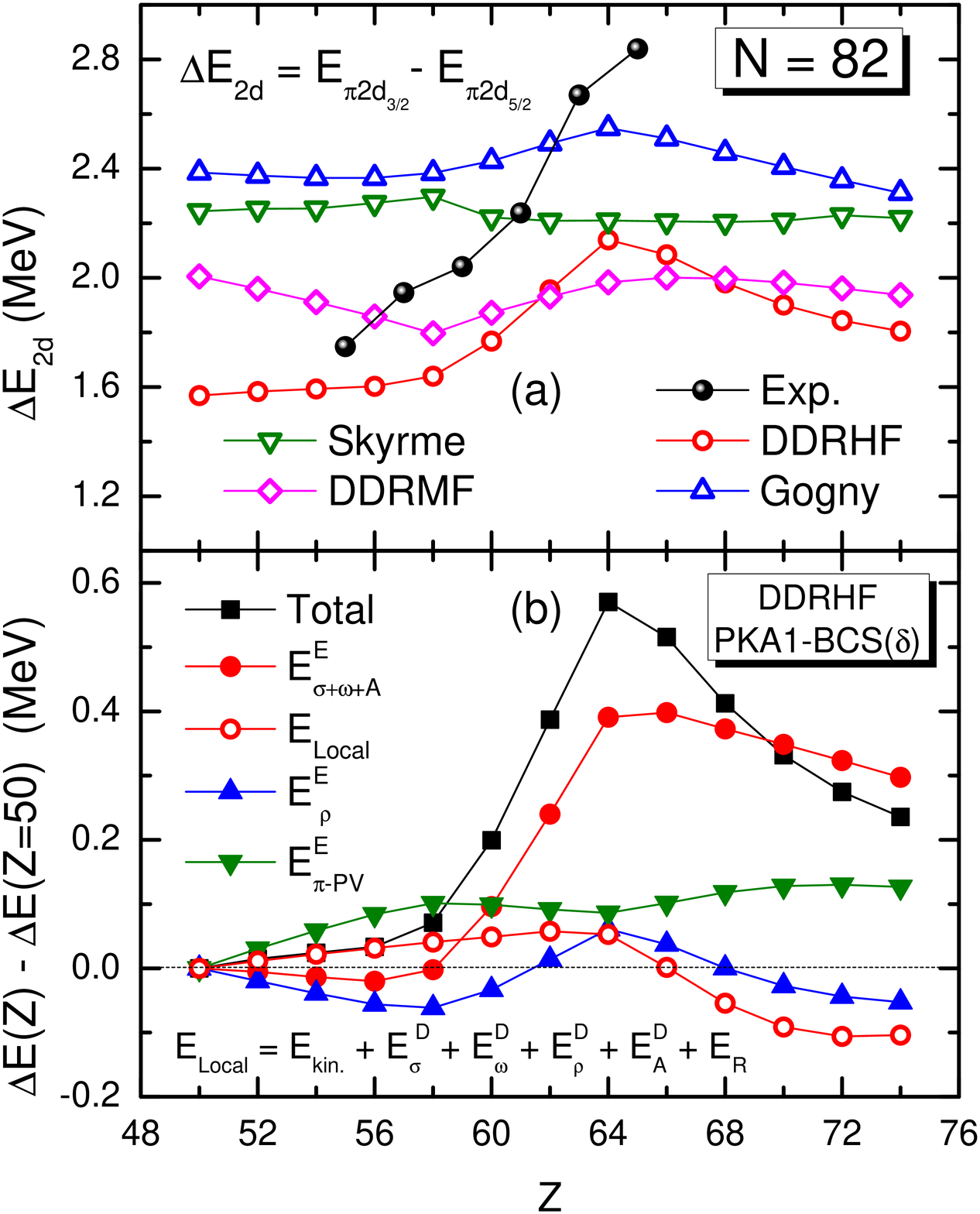}
\caption{(Color online) (a) Spin-orbit splittings $\Delta E_{2d}$ of proton $\pi2d$ states
extracted from the calculations of DDRHF with PKA1 \cite{Long:2007}, Gogny HFB with D1S
\cite{Berger84}, Skyrme HF with SLy4 and RMF with PKDD \cite{Long04}, in comparison with
experimental data \cite{Nagai:1981}; (b) Detailed contributions of $\Delta E_{2d}$ from different
channels given by the DDRHF calculations with PKA1. }\label{fig:Z64_rhfP}
\end{figure}

For the spin-orbit splitting of $\pi2d$ states, namely the $Z=64$ sub-shell in Fig.
\ref{fig:LevP_Gd146}, the isotonic evolutions given by these four models are also shown in Fig.
\ref{fig:Z64_rhfP}a, in comparison with the data \cite{Nagai:1981} which represents remarkable
enhancement ($\sim$1MeV) towards $Z=65$. Among the theoretical results, DDRHF calculations present the strongest enhancement ($\sim$0.6MeV) as compared to Gogny and DDRMF results ($\sim$0.2MeV).
Much different from experimental trend, the Skyrme calculations give nearly constant spin-orbit
splitting even slightly quenching from $Z=58$ to 64 and DDRMF model gives the kink at $Z=58$. Thus, the non-local mean fields in DDRHF and Gogny models may play the important role to reproduce the
general trend of the isotonic dependence of the spin-orbit splittings. More specifically Fig.
\ref{fig:Z64_rhfP}b shows detailed contributions of the spin-orbit splitting $\Delta E_{2d}$ from
different channels given by the DDRHF calculations. In Fig. \ref{fig:Z64_rhfP}b it is clearly seen that the enhancement of the spin orbital splitting is mainly determined by the non-local Fock terms whereas very little effect is induced by the local term, i.e., $E_{\text{Local}} = E_\kin +
E_\sigma^\RD + E_\omega^\RD + E_\rho^\RD + E_A^\RD +E_R$.

Among the Fock channels, the isoscalar $\sigma$ and $\omega$ mesons together with the Coulomb
field, i.e., $ E_{\sigma+\omega+A}^\RE = E_\sigma^\RE+E_\omega^\RE + E_A^\RE$, play the dominant
role in determining the enhancement of spin-orbit effects towards $Z=64$. Compared to strong
isoscalar exchange terms, the isovector ones from $\pi$ and $\rho$ nucleon couplings present small
effects. In fact the proton shell evolution along the isotonic chain, consistent with proton
configuration, is mainly determined by the proton-proton interaction, among which the isoscalar
exchange terms contribute much stronger effects than the isovector ones. As compared to the
$T=0$ tensor force \cite{Otsuka:2005, Colo:2007, Lesinski:2007, Nakada:2008, Long:2008}, the tensor effects here ($T=1$) are fairly weak.  In Fig. \ref{fig:Z64_rhfP}b distinct cancelation between
$\pi$ and $\rho$ (mainly tensor $\rho$) contributions is found from $Z=50$ to $58$, where the
valence protons will gradually occupy the $\pi1g_{7/2}$ state. Because of the tensor feature
\cite{Long:2008}, the spin-orbital splittings are therefore enlarged by the one-pion exchange
whereas the tensor $\rho$ presents the opposite effects. As shown in Fig. \ref{fig:Z64_rhfP}b, such cancelation are always found along the isotonic chain of $N=82$. The effects from the isovector
channels are then strongly weakened. The non-local mean field, mainly the isoscalar exchange terms, therefore becomes specially significant to reproduce the enhancement of spin-orbit effect in
$\pi2d$ states of the $N=82$ isotones.

In addition if comparing the pseudo-spin and spin orbital splittings in Fig.
\ref{fig:Z64-PSS}a and Fig. \ref{fig:Z64_rhfP}a, one can find that the data represent evident
consistency between the systematics of pseudo-spin and spin orbital splittings, i.e., stronger
spin-orbit effect accompanied by better preserved pseudo-spin symmetry. Among the theoretical
calculations, such consistence between pseudo-spin degenerations and spin-orbit splitting is
reproduced completely by DDRHF and partially by Gogny.

\begin{figure}[t]
\includegraphics[width=0.48\textwidth]{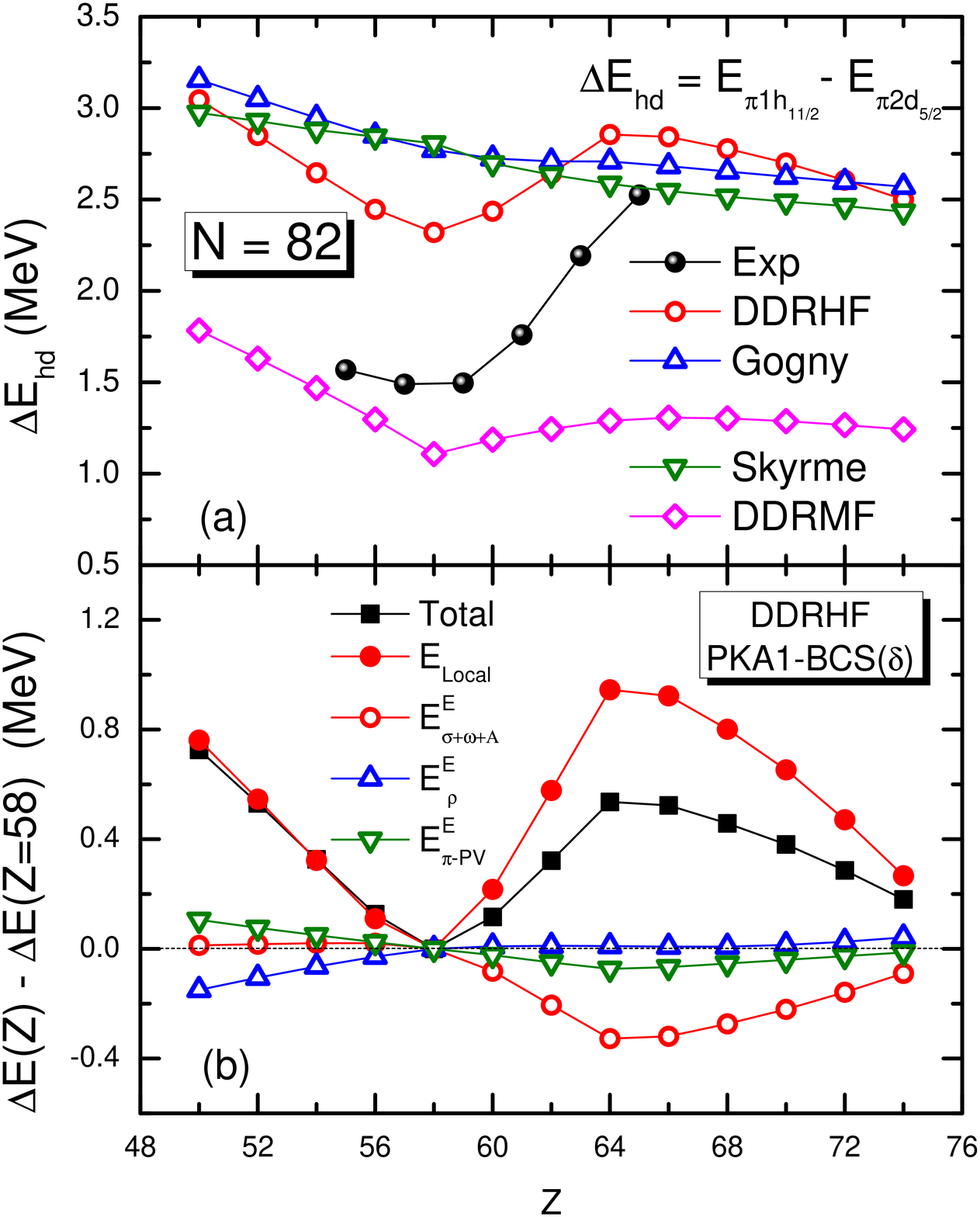}
\caption{(Color online) (a) Energy difference $\Delta E_{hd}$ between proton states $\pi1h_{11/2}$
and $\pi2d_{5/2}$ extracted from the calculations of DDRHF with PKA1 \cite{Long:2007}, Gogny HFB
with D1S \cite{Berger84}, Skyrme HF with SLy4 and RMF with PKDD \cite{Long04}, in comparison with
experimental data \cite{Nagai:1981}; (b) Detailed contributions of $\Delta E_{hd}$ from different
channels given by the DDRHF calculations with PKA1.}\label{fig:H11_RHFP}
\end{figure}

Experimentally the $Z=64$ sub-shell is determined by the energy difference $\Delta E_{hd}$ between
the $\pi2d_{5/2}$ and $\pi1h_{11/2}$ single-particle orbits \cite{Nagai:1981}. In Fig.
\ref{fig:H11_RHFP}a are shown the energy differences $\Delta E_{hd}$ along the $N=82$ isotonic
chain. For comparison are also shown the experimental data extracted from Nagai et al. in Ref.
\cite{Nagai:1981}, which is consistent with the experimental systematics along the $N=83$ chain
extracted by Odahara et al., from the analysis of high spin isomers \cite{Odahara:1997}. Among the
theoretical calculations, DDRHF with PKA1 presents similar substantial enhancement ($\sim$0.5MeV)
as the experimental one ($\sim$1MeV) toward Z=64 sub-shell. Fig. \ref{fig:H11_RHFP}b presents the
contributions of $\Delta E_{hd}$ from different channels, given by the DDRHF calculations with
PKA1. From Fig. \ref{fig:H11_RHFP}b it is found that the systematics of the energy difference
$\Delta E_{hd}$ is determined by the local term $E_{\text{Local}}$ while the isoscalar non-local terms $E_{\sigma+\omega+ A}^\RE$ present opposite contributions. Here the contributions from the isovector exchange terms become negligible. One may notice that the energy difference $\Delta E_{hd}$ is given by two spin-up states, which has different physical mechanism from the spin-orbit splitting of $\pi2d$ states. From this point view, one can understand different effects of the exchange terms on the spin-orbit splittings and energy difference $\Delta E_{hd}$ as shown in Fig. \ref{fig:Z64_rhfP}b and Fig. \ref{fig:H11_RHFP}b, respectively.

In conclusion, the $Z=64$ sub-shell evolution along the isotonic chain of $N=82$ is investigated
by relativistic and non-relativistic mean-field approaches. Mean fields in DDRMF and Skyrme HF
are all local in the coordinate space, while those in DDRHF and Gogny HFB have non-local exchange
contributions. The systematic trend of spin-orbit and pseudo-spin-orbit splitting of $N=82$ isotones are also studied. The pseudo-spin symmetry is well reproduced by DDRHF, Gogny-HFB and Skyrme HF. In contrast, DDRMF produces too large splitting, leading to an unphysical shell closure of $Z=58$. DDRHF and Gogny can also acount for the $Z$-dependence of the pseudo-spin splitting. The empirical large increase of the spin-orbit splitting between $\pi2d_{3/2}$ and $\pi2d_{5/2}$ states towards Z=64 sub-shell is reproduced by the theoretical results calculated by DDRHF, and also by Gogny but less extent. It is found in DDRHF that the non-local effects due to the isoscalar $\sigma$ and $\omega$ exchange terms play an essential role in determining the enhancements of the spin-orbit splitting of $\pi2d$ states, which creates properly the $Z=64$ sub-shell closure at the same time. For the spin-orbit splitting, the $\pi$ and $\rho$ tensor contributions cancel each other and the net contribution becomes rather small. Although there exists some discrepancy on the order of states, the enhancement of the experimental sub-shell gap towards $Z=64$, namely the energy difference between the $\pi2d_{5/2}$ and $\pi1h_{11/2}$ states, is reproduced by the DDRHF model for which the local terms of the scalar and vector mesons are responsible.

We would like to thank A. Odahara and M. Gono for providing us experimental information on the
sub-shell structure $Z=64$. This work is partially supported by the Grant-in-Aid for Scientific Research of JSPS (Nos. 20540277 and 31340073), the National Natural Science Foundation of China under Grant No. 10775004, and the Major State Basic Research Development Program Under Contract No. 2007CB815000.

%\bibliographystyle{elsarticle-num}
%\bibliography{D:/ÂÛÎIJݸå/rhffr}

\begin{thebibliography}{10}
\expandafter\ifx\csname url\endcsname\relax
  \def\url#1{\texttt{#1}}\fi
\expandafter\ifx\csname urlprefix\endcsname\relax\def\urlprefix{URL }\fi
\expandafter\ifx\csname href\endcsname\relax
  \def\href#1#2{#2} \def\path#1{#1}\fi

\bibitem{Tan:1985}
I.~Tanihata, H.~Hamagaki, O.~Hashimoto, Y.~Shida, N.~Yoshikawa, K.~Sugimoto,
  O.~Yamakawa, T.~Kobayashi, N.~Takahashi, {M}easurements of {I}nteraction
  {C}ross {S}ections and {N}uclear {R}adii in the {L}ight p-{S}hell {R}egion,
  Phys. Rev. Lett. 55 (1985) 2676--2679.

\bibitem{Meng:1996}
J.~Meng, P.~Ring, {R}elativistic {H}artree-{B}ogoliubov {D}escription of the
  {N}eutron {H}alo in $^{11}${L}i, Phys. Rev. Lett. 77 (1996) 3963--3966.

\bibitem{Nagai:1981}
Y.~Nagai, J.~Styczen, M.~Piiparinen, P.~Kleinheinz, {P}roton single-particle
  states above $z=64$, Phys. Rev. Lett. 47 (1981) 1259.

\bibitem{Odahara:1997}
A.~Odahara, Y.~Gono, S.~Mitarai, T.~Morikawa, T.~Shizuma, M.~Kidera,
  M.~Shibata, T.~Kishida, E.~Ideguchi, K.~Morita, A.~Yoshida, H.~Kumagai,
  Y.~Zhang, A.~Ferragut, T.~Murakami, M.~Oshima, H.~Iimura, M.~Shibata,
  S.~Hamada, H.~Kusakari, M.~Sugawara, M.~Ogawa, M.~Nakajima, B.~Min, J.~Kim,
  S.~Chae, H.~Sagawa, {H}igh-spin isomer and level structure of $^{145}$sm,
  Nucl. Phys. A 620 (1997) 363--384.

\bibitem{Walecka:1974}
J.~D. Walecka, {A} theory of highly condensed matter, Ann. Phys. (N.Y.) 83
  (1974) 491--529.

\bibitem{Serot:1986}
B.~D. Serot, J.~D. Walecka, {T}he relativistic nuclear many-body problem, Adv.
  Nucl. Phys. 16 (1986) 1--327.

\bibitem{Geng:2006}
L.-S. Geng, J.~Meng, H.~Toki, W.-H. Long, G.~Shen, Spurious shell closures in
  the relativistic mean field model, Chin. Phys. Lett. 23 (2006) 1139--1141.

\bibitem{Long:2007}
W.~H. Long, H.~Sagawa, N.~V. Giai, J.~Meng, Shell structure and $\rho$-tensor
  correlations in density dependent relativistic {H}artree-{F}ock theory, Phys.
  Rev. C 76 (2007) 034314.

\bibitem{Long:2006}
W.~H. Long, N.~V. Giai, J.~Meng, {D}ensity-dependent relativistic
  {H}artree-{F}ock approach, Phys. Lett. B 640 (2006) 150--154.

\bibitem{Long:2006PS}
W.~H. Long, H.~Sagawa, J.~Meng, N.~V. Giai, {P}seudo-spin symmetry in
  density-dependent relativistic {H}artree-{F}ock theory, Phys. Lett. B 639
  (2006) 242--247.

\bibitem{Long:2008}
W.~H. Long, H.~Sagawa, J.~Meng, N.~V. Giai, Evolution of nuclear shell
  structure due to the pion exchange potential, Europhysics Letters 82 (2008)
  12001.

\bibitem{Liang:2008}
H.~Z. Liang, N.~V. Giai, J.~Meng, Spin-isospin resonances: A self-consistent
  covariant description, Phys. Rev. Lett. 101 (2008) 122502.

\bibitem{BYSun:2008}
B.~Y. Sun, W.~H. Long, J.~Meng, U.~Lombardo, {N}eutron star properties in
  density-dependent relativistic {H}artree-{F}ock theory, Phys. Rev. C 78
  (2008) 065805.

\bibitem{Long:2009}
W.~H. Long, P.~Ring, N.~V. Giai, J.~Meng, Relativistic
  {H}artree-{F}ock-{B}ogoliubov theory with density-dependent meson-nucleon
  couplings\href {http://arxiv.org/abs/nucl-th/0812.1103}
  {\path{arXiv:nucl-th/0812.1103}}.

\bibitem{Reinhard:1989}
P.-G. Reinhard, The relativistic mean-field description of nuclei and nuclear
  dynamics, Reports on Progress in Physics 52 (1989) 439--514.

\bibitem{Ring:1996}
P.~Ring, Relativistic mean field theory in finite nuclei, Prog. Part. Nucl.
  Phys. 37 (1996) 193--263.

\bibitem{Meng:2006}
J.~Meng, H.~Toki, S.~G. Zhou, S.~Q. Zhang, W.~H. Long, L.~S. Geng,
  {R}elativistic {C}ontinuum {H}artree-{B}ogoliubov theory for the ground state
  properties of exotic nuclei, Prog. Part. Nucl. Phys. 57 (2006) 470--563.

\bibitem{Vretenar:2005}
D.~Vretenar, A.~V. Afanasjev, G.~A. Lalazissis, P.~Ring, {R}elativistic
  {H}artree-{B}ogoliubov theory: static and dynamic aspects of exotic nuclear
  structure, Phys. Rep. 409 (2005) 101--259.

\bibitem{Bouyssy:1987}
A.~Bouyssy, J.~F. Mathiot, N.~V. Giai, S.~Marcos, {R}elativistic description of
  nuclear systems in the {H}artree-{F}ock approximation, Phys. Rev. C 36 (1987)
  380--401.

\bibitem{Berger84}
J.~F. Berger, M.~Girod, D.~Gogny, {M}icroscopic analysis of collective dynamics
  in low energy fission, Nucl. Phys. A 428 (1984) 23--36.

\bibitem{Chabanat:1998}
E.~Chabanat, P.~Bonche, P.~Haensel, J.~Meyer, R.~Schaeffer, {A} {S}kyrme
  parametrization from subnuclear to neutron star densities part {II. N}uclei
  far from stabilities, Nucl. Phys. A 635 (1998) 231--256.

\bibitem{Long04}
W.~H. Long, J.~Meng, N.~V. Giai, S.-G. Zhou, {N}ew effective interactions in
  {RMF} theory with non-linear terms and density-dependent meson-nucleon
  coupling, Phys. Rev. C 69 (2004) 034319.

\bibitem{Dobaczewski:1996}
J.~Dobaczewski, W.~Nazarewicz, T.~R. Werner, J.~F. Berger, C.~R. Chinn,
  J.~Decharg$\acute{\rm e}$, Mean-field description of ground-state properties
  of drip-line nuclei: Pairing and continuum effects, Phys. Rev. C 53 (1996)
  2809--2840.

\bibitem{Otsuka:2005}
T.~Otsuka, T.~Suzuki, R.~Fujimoto, H.~Grawe, Y.~Akaishi, Evolution of nuclear
  shells due to the tensor force, Phys. Rev. Lett. 95 (2005) 232502.

\bibitem{Colo:2007}
G.~Col$\grave{\rm o}$, H.~Sagawa, S.~Fracasso, P.~F. Bortignon, Spin-orbit
  splitting and the tensor component of the skyrme interaction, Phys. Lett. B
  646 (2007) 227--231.

\bibitem{Lesinski:2007}
T.~Lesinski, M.~Bender, K.~Bennaceur, T.~Duguet, J.~Meyer, Tensor part of the
  skyrme energy density functional: Spherical nuclei, Phys. Rev. C 76 (2007)
  014312.

\bibitem{Nakada:2008}
H.~Nakada, Mean-field approach to nuclear structure with semi-realistic
  nucleon-nucleon interactions, Phys. Rev. C 78 (2008) 054301.

\end{thebibliography}
%\end{document}

\end{document}